# RNA structure prediction: progress and perspective[*]

Shi Ya-Zhou(时亚洲), Wu Yuan-Yan(吴园燕), Wang Feng-Hua(王凤华), and Tan Zhi-Jie(谭志杰)[†]

Department of Physics and Key Laboratory of Artificial Micro- and Nano-structures of Ministry of Education, School of Physics and Technology, Wuhan University, Wuhan 430072, China

## ABSTRACT

Many recent exciting discoveries have revealed the versatility of RNAs and their importance in a variety of cellular functions which are strongly coupled to RNA structures. To understand the functions of RNAs, some structure prediction models have been developed in recent years. In this review, the progress in computational models for RNA structure prediction is introduced and the distinguishing features of many outstanding algorithms are discussed, emphasizing three dimensional (3D) structure prediction. A promising coarse-grained model for predicting RNA 3D structure, stability and salt effect is also introduced briefly. Finally, we discuss the major challenges in the RNA 3D structure modeling.

[*] Project supported by the National Science Foundation of China grants (11074191, 11175132 and 11374234), the National Key Scientific Program (973)-Nanoscience and Nanotechnology (No. 2011CB933600), and the Program for New Century Excellent Talents (NCET 08-0408).

[†] Corresponding author. Email: zjtan@whu.edu.cn

# 1. Introduction

It is well known that the primary role of RNAs is to convert the genetic information stored in DNA into proteins. Rapid methods for sequencing DNA have revealed that over 75% of human genomic DNA is transcribed into RNA even though only about 5% of the DNA codes are translated into proteins.[1] Within the past few years, many additional functions of non-coding RNAs, i.e., functional but not in the sense of being translated into protein, have been discovered, which indicates that RNAs play more important roles in the life process than that stipulated by the traditional consensus in molecular biology.[1,2] Besides the traditional non-coding RNAs such as the ribosomal RNAs and transfer RNAs, other non-coding RNAs have been discovered, including ribozymes, small interface RNAs and riboswitches, which are known, respectively, to catalyze mRNA splicing, control protein synthesis and regulate transcription and translation.[3-7]

To become functional, RNAs generally adopt compact tertiary structures. Understanding these structural features not only gives insights into structure-function relationship but also aids drug design.[8] Until now, the experimental methods for measuring RNA three-dimension (3D) structure include X-ray crystallography and NMR spectroscopy, both of which have been successfully employed to obtain high-resolution RNA 3D structures. However, due to the flexible nature of RNAs and the high cost of the methods, the number of RNA structures in the database is much less than that of RNA sequences. Therefore, computational modeling becomes very important to obtain RNA structures.[9-12]

The RNA folding process is generally hierarchical, i.e., helical elements in a secondary structure are formed first and then the compact tertiary structures are folded with the associations of RNA motifs, the formation of tertiary contacts and possible minor rearrangement of secondary segments.[8,13,14] Due to the strong base-pairing and base-stacking interactions, RNA secondary structures usually are relatively stable. Therefore, RNA folding can be generally divided into two steps: RNA secondary structure folding and RNA tertiary structure folding. In the literatures, RNA structure modeling is generally classified into two levels: 2-dimensional secondary structure modeling and 3D structure modeling.

In this review, we will focus on the recent advances in RNA structure prediction, and the main text is organized as follows. In Section 2, we will give a brief overview on the predictive algorithms for 2D RNA secondary structures prediction. In Section 3, we will give a detailed introduction for the recent progress in modeling RNA 3D structure, including the knowledge-based models and physics-based models. At the end of the Section 3, we will briefly introduce a predictive coarse-grained (CG) model, newly developed for predicting RNA 3D structure, stability and salt effects. Finally, we will discuss the challenges in modeling RNA 3D structure.

## 2. RNA secondary structure prediction

RNA folding process is generally hierarchical, which means that local interactions occur first and are energetically stronger than tertiary interactions.[8,13,14] Therefore, the RNA secondary structure provides a scaffold to its native 3D structure, and it can be predicted without the knowledge of tertiary interactions. Over the past three decades, there are many computational models have been developed for predicting RNA secondary structure, and these have been reviewed elsewhere.[13,15-17] The major models are listed in Table 1 along with the corresponding websites and references. These computational models can be roughly classified into two categories: comparative sequence-based methods and free energy-based methods. In the following, we will give a brief introduction on the models.

### 2.1. Comparative sequence-based methods

Assuming that RNAs with the similar sequences in different species have the similar structures, the secondary structures of RNAs can be predicted by comparative sequence analysis,. This logic was first used to predict the secondary structure of tRNA correctly based on the expert knowledge of RNAs.[9,18]

Since the incipient manual comparative sequence analysis methods highly depend on the knowledge of users and are time-consuming, several automatic algorithms[19-25] that use multiple sequences were recently developed to predict RNA secondary structures, such as RNAalifold[20], Pfold/PPfold[21,22], TurboFold[23], RNAforester[24] and MARNA[25]. These algorithms have been benchmarked for prediction speed and accuracy with the same data set[26] and have been reviewed elsewhere[16]. Here, we only give a brief overview.

RNAalifold predicts the consensus structure for a set of sequences from a given alignment with the use of sequence covariations and thermodynamic parameters.[20] Pfold couples a stochastic context-free grammar to a phylogenetic analysis with a high accuracy in predictions,[21] and its multithreaded version, PPfold can predict the structures for large RNA alignments at practical timescales with the phylogenetic calculations and the inside-outside algorithm.[22] An iterative method, TurboFold can predict secondary structures for multiple RNA sequences with the free energy model and the information from the comparative analysis between sequences.[23] RNAforester[24] and MARNA[25] are the algorithms that firstly fold the RNA sequences with the free energy-based methods and then align the predicted structures with special tools.

## 2.2. Free energy-based methods

The most popular methods are to predict RNA secondary structures from single sequence by calculating the free energy, which is based on the thermodynamic hypothesis that the conformation with the lowest free energy is the native structure.[27]

2.2.1. Free energy minimization

Based on an empirical nearest-neighbor model and thermodynamic parameters obtained from experiments,[28] Mfold[29] is a very important algorithm for searching the RNA secondary structure with minimum free energy from a single sequence with the use of the dynamic programming algorithm. Mfold can predict secondary structures for RNAs with accuracy of ~70%.[29] Similarly, RNAstructure[30] and RNAfold[31] were developed for predicting the structures with minimum free energies as well as calculating the equilibrium partition functions and base pairing probabilities.

However, because of the assumption of free energy additivity and the absence of measured thermodynamic parameters of non-canonical base pairs and sequence-dependent loops, the free energy computed from the nearest-neighbor model for a conformation is not accurate.[32] Consequently, some advanced methods have introduced statistical data or abstract shapes of known structures to more accurately sample RNA secondary structures.[33-35] The Sfold is a typical statistical algorithm developed by Ding and Lawrence.[34] With a partition function calculation, Sfold can sample the optimal secondary structure according to Boltzmann probability with a stochastic dynamic programming algorithm.[34] RNAshapes uses the concept of abstract shapes to calculate cumulative probabilities of structures belonging to the identified shapes, and integrates the shapes well with dynamic programming algorithms to efficiently predict RNA secondary structures.[35]

2.2.2. Suboptimal structure prediction

Some RNA sequences may adopt secondary structures that are partially determined by folding kinetics or may have more than one conformation.[36] The simplest method is to explicitly generate all possible structures from sequence, based on the thermodynamic parameters, and to further give possible folding pathways with the master equation method.[36] However, it is generally impossible to enumerate all possible structures for large RNAs due to the huge number of possible structures for a sequence.[15] To predict the multiple kinetically-stable structures for large RNAs would require the approximations at different element-levels.[37]

Recently, some automated procedures for efficiently generating a diverse set of suboptimal structures for large RNAs were proposed.[38,39] The Kinwalker introduced a heuristic approach for folding kinetics of large RNAs by constructing secondary structures through stepwise combination of locally optimal substructures, which can be calculated by dynamic programming.[38] Another method of the MPGAfold can predict possible folding pathways and the structures of functional intermediates with the use of a massively parallel genetic algorithm.[39]

2.2.3. Pseudoknot structure prediction

An RNA pseudoknot, which is minimally composed of two helical segments connected by single-stranded regions or loops, plays diverse fundamental roles in the control of viral replication, in structural organization of complex RNAs and in the self-cleaving ribozyme catalysis.[40] Since there are crosslinking of pseudoknot loops and thermodynamic parameters for pseudoknot loops have not been experimentally derived, to accurately predict secondary structures of RNA pseudoknots is still challenging, especially for the complex non-H-type pseudoknots.[17,40-47]

To address the problem, Rivas and Eddy introduced a few pseudoknot-specifc parameters to fold optimal pseudoknotted RNAs (100~200 nt) with a dynamic programming algorithm.[41] The ILM model uses a combination of thermodynamic and covariance information to predict RNA secondary structures including pseudoknots through an iterated loop matching algorithm.[42] As a heuristic algorithm, HotKnots predicts secondary structures of pseudoknots by assembling the energetically favorable substructures determined from the free energy model.[43] Similarly, FlexStem constructs RNA secondary structures with pseudoknots by adding maximal stems according to the free energy model.[44] Kinefold predicts the structures of pseudoknots with topological and geometrical constraints through a long-time-scale RNA folding simulation,[45] which follows the stochastic closing and opening of individual RNA helices. In addition, the thermodynamic parameters of pseudoknots loop can also be derived from a lattice model[40,47,48] and based on which, the Vfold predicts the free energy landscapes of pseudoknots.[40,48]

## 3. RNA 3D structure prediction

Although the secondary structure provides the blueprint of an RNA molecule, the knowledge of RNA 3D structure is still indispensably for understanding its function comprehensively. For initial 3D structure modeling, the 3D structures of some common RNA molecules such as tRNAs,[9] group I introns[10] and RNase P[11] have been successfully built by RNA structure experts. In recent years, a variety of computational models have been developed for predicting RNA 3D structures,[13,14,49-51] which have been tabulated in Table 2. These models can be roughly classified into two categories, based on whether they are knowledge-based or physics-based.

### 3.1. Knowledge-based modeling

With the increasing number of experimentally determined structures in database, RNA 3D structures can be predicted by the assembly of known motifs or sequence alignment. Knowledge-based modeling mainly includes graphics-based modeling and homology-based modeling.

### 3.1.1. Graphics-based methods

Graphics-based modeling generally provides a graphical interface and allows users to construct RNA 3D structures by manipulating or assembling RNA segments.[52-56] The major graphics-based algorithms will be introduced as follows.

*MANIP*. As an interactive tool, the MANIP allows the assembly of known 3D motifs into a complete RNA structure on the computer screen by users from the corresponding secondary structure predicted by comparative sequence analysis.[52] Although the MANIP is not an automatic procedure, it constitutes a quick and easy way to build 3D structures of RNAs especially large-size RNAs, e.g., the RNase P RNA and the group I introns. In addition, multiple connection and base-pair tables that explicitly contain topological information of RNAs, can be a reference to the precise modeling of interactions between RNAs.[52]

*ERNA-3D*. To generate RNA 3D structures from sequences and secondary structures, the ERNA-3D provides a graphical interface for users to freely position the A-form helices and to directly pull the single inter-helical strands.[53] The main advantage of the algorithm is that there is almost no limitation to the size of RNAs. ERNA-3D has been used to successfully generate the 3D structures of mRNA, tRNAs and rRNAs, including the 16 S rRNA, 23 S rRNA and 5 S rRNA.[53]

*RNA2D3D*. The RNA2D3D can rapidly predict rough 3D structures for large RNAs, e.g., ribozymes, viral kissing loops and variety of RNA nanostructures, based on their secondary structures.[54] However, the manual manipulation is required to generate satisfactory 3D structures through a graphical interface with the special tools such as the compacting, stem-stacking, segment-positioning and energy-refinement.[54]

*S2S/Assemble*. The S2S/Assemble is an interactive graphical algorithm in which users can easily display, manipulate and interconnect RNA data from sequence to structure, as well as analyzing and building RNA 3D architectures.[55,56] Since all interactions including base-pairing and base-stacking have to be annotated manually, it is difficult to perform a high-throughput structure prediction with this algorithm.

Although the graphics-based methods introduced above can be used to build 3D structures for large RNAs with hundreds of nucleotides rapidly and intuitively, because they are manual tools, they need users to set up and refine the RNA structure models according to specific principles using the tools included in the software packages. Thus, in order to build plausible structures, it is necessary for users to have intimate knowledge of RNA structures.

3.1.2. Homology-based modeling

Since the 3D structure of a macromolecule evolves much more slowly than its sequence, evolutionarily related macromolecules usually retain similar 3D structure despite the divergence on the sequence level. Based on this fact, the 3D structures of a macromolecule can be built through aligning the sequence of the target molecule to the structures of template molecules.[56] Homology-based modeling, also called template-based modeling or comparative modeling, has been rather successful in predicting 3D structure of proteins.[57,58] In analogy to the homology-based modeling for proteins, several algorithms[59-63] such as ModeRNA,[59] and RNABuilder[60] have been developed for building RNA 3D structures. In addition, the homology-based modeling has been generalized to the fragment-assembly methods, such as 3dRNA[61,62] and RNAComposer[63].

*ModeRNA*. ModeRNA allows both simplistic structure prediction from a set of templates/alignments and user-controlled manipulations of structures, e.g., fragment assembly.[59] Compared with other modeling algorithms, ModeRNA can recognize and model post-transcriptionally modified nucleosides. It should be noted that although ModeRNA is not a graphics-based tool, it still requires users to supply the alignment between the template RNAs and the target RNA, and to specify the base pairs between the inserted fragment and the rest of the RNA.

*RNABuilder*. RNABuilder, now known as the MMB, is another method for comparative modeling of RNA structures by using the distance and torsion angles from the aligned regions of the template as modeling restraints.[60] The distinguishing characteristics of RNABuilder is that it introduces a CG force field including base-pairing, base-stacking, tertiary interactions, and excluded-volume interactions, to bring the interacting bases into the base pairing geometry. The software package is written with the internal coordinate mechanics library, so the CG force field can be applied to the base rather than to individual atoms. Although all forces employed in RNABuilder are user-specified, it provides a new framework for predicting RNA 3D structures combining with experimental constraints and users' hypotheses. In addition, RNABuilder has been used to successfully build the entire 200-nt Azoarcus ribozyme structure with structural information from two template structures.[60]

*3dRNA*. 3dRNA is a fast and automated algorithm for building RNA 3D structure by assembling the A-form helices and various loops whose structures are extracted from the known structures in database.[61,62] For 300 tested RNAs including duplexes, hairpins, pseudoknots and junctions, 3dRNA predicts reliable 3D structures based on their secondary structures. In addition, as a web server, 3dRNA can be used freely online, and with the sequence and secondary structure as input, one can obtain the predicted 3D structure quickly.[61]

*RNAComposer*. RNAComposer is another user-friendly web server and it can predict 3D structures for large RNAs from their secondary structures based on the RNA FRABASE database.[63] The RNA FRABASE database can be regarded as a dictionary which relates the RNA secondary structure with known tertiary structure fragments. In RNAComposer, the secondary structure inputted by the user is firstly fragmented into the elements such as stems, loops and single strands, and afterwards, the corresponding tertiary structure elements are automatically searched out from the RNA FRABASE database and assembled into complete 3D structures.

As introduced above, the major advantage of homology-based modeling is that there is generally no limitation on size of RNAs to be modeled. However, the quality of the predicted structures is dependent on the quality of the sequence alignments, template structures and user-defined secondary structures. Although the number of known RNA structures stored in the PDB/NDB database is increasing rapidly, it still may not be easy to find proper template RNAs for a particular target RNA. Moreover, due to the high flexibility of RNAs, their structures generally change with the solution environments such as temperature, ion conditions[64] and existence of other ligands or macromolecules. In addition, the development of a good alignment for RNAs with complex structures usually requires laborious manual preparation based on the established expertise with regard to the most relevant RNA families. Thus, homology-based modeling is not always effective.

## 3.2. Physics-based modeling

Physics-based approaches are based on biophysical principles, which simulate the folding process in search of the conformation with minimal free energy. Since a full-atom structure modeling for an RNA generally involves many degrees of freedom and consequently huge computational complexity, several CG predictive models have also been proposed at different resolution levels with physical simplifications.

### 3.2.1. All-atomistic model

Until now, the all-atomistic molecular dynamics is popular for macromolecule modeling and can give a view of the actual motion of atoms with physics-based atomic force fields such as AMBER[65,66] and CHARMM[67]. However, due to the many degrees of freedom, it is still difficult to fold RNA 3D structures with these all-atomistic force fields even with modern computational facilities. Consequently, the models with the all-atomistic fragments assembling have been developed based on the known fragments or secondary structures,[68-76] such as MC-fold/Mc-Sym pipeline[68], FARNA/FARFAR[69,70], RSIM[71] and RNAnbds[73,74].

*MC-fold/MC-sym.* Since secondary structures would provide enough structural constraints to automated 3D building, the MC-Fold/MC-Sym pipeline infers RNA secondary structure from sequence data and then assembles a series of 3D structures based on their secondary structures.[68] Unlike the thermodynamics approaches such as Mfold[29], the MC-Fold can predict RNA secondary structures including canonical and noncanonical base pairs with the use of a knowledge-based scoring function based on the database of NCMs (nucleotide cycle motifs). The NCMs, which are circularly connected by covalent, pairing or stacking interactions, was built from an analysis of the X-ray crystallographic structures. The fragment insertion simulation is performed by MC-Sym with the 3D NCMs and the Las Vegas algorithm. MC-fold/MC-Sym pipeline has been validated by building 3D structures of precursor microRNA and a new 3D structure of the human immunodeficiency virus (HIV1) *cis*-acting -1 frame-shifting segment.[68]

*FARNA/FARFAR*. Das and Baker explored a fully automated and energy-based approach of FARNA to predict RNA 3D structure.[69] With the use of Monte Carlo algorithm and a simplified knowledge-based energy function that favors base pairing and stacking geometries, FARNA assembles trinucleotide fragments extracted from the ribosome crystal structure into an all-atomistic structure corresponding to the given sequence. The CG base-pairing potential used in FARNA is based on the statistical analysis of the bases in the ribosome and can account for not only Watson-Crick base pairs but also the interactions with the Hoogsteen and sugar edges. The FARFAR introduces a high-resolution refinement phase into FARNA to predict and design non-canonical small RNA structures with atomic accuracy.[70]

*RSIM*. As an improved method of FARNA,[69] the RSIM predicts RNA 3D structures from secondary structure constraints using Monte Carlo algorithm with closed moves.[71] The important feature of RSIM is that it allows visualization of the predicted RNA conformational space as a graph which provides insights into the possible RNA dynamics to further identify functional RNA conformations. However, the current implementation of RSIM is limited to non-pseudoknotted RNAs and requires manual intervention for complicated branched structures.[71]

*RNAnbds*. The RNAnbds is designed to predict RNA 3D structures by fragment assembly based on the statistical configurations of bases according to their sequence/spatial neighbors in databases.[73,74] Combined with a statistical potential including base-pairing and base-stacking and a sequential Monte Carlo method, RNAnbds gives reliable prediction for short fragments (<15 nt) especially loops with RMSDs <4Å between the predicted and the experimental fragments.

### 3.2.2. Coarse-grained model

Another way to reduce the computational cost is to reduce the number of particles by treating a group of functional atoms as a single bead.[77-79] The bead may represent only a few atoms or a large group of atoms depending on the resolution of the model; see Fig. 1. After the one-bead RNA model firstly developed by Malhotra and Harvey,[80] there are many CG models have been implemented to predict RNA 3D structures[81-90] or to model the interactions between RNAs and other molecules[91,92], such as YUP[81], NAST[82], iFold[83,84], Vfold[85,86], etc. (see Table 2).

*YUP*. The YUP is a very flexible molecular mechanics algorithm for CG and multi-scaled modeling.[81] Based on the associated energy potentials and the methods such as Monte Carlo, molecular dynamics and energy minimization, the algorithm has been used to model RNA structures as well as structures of protein and DNA. In YUP, to model large RNAs, one nucleotide is replaced by one pseudoatom placed at the center of phosphorus atom, which effectively reduces the computational cost. Although YUP requires users to provide the secondary structure of RNAs and the information for the force field, it is a adaptive package for automatic CG modeling of RNAs.[81]

*NAST*. Similar to YUP[81], the NAST is another one-bead model, in which the C3' atom in a nucleotide is chosen to represent the entire nucleotide.[82] With an RNA-specific knowledge-based potential and a simple molecular dynamics algorithm, NAST can sample conformations that satisfy a given set of secondary structure and tertiary contact constraints. One advantage of NAST is its ability to incorporate experimental data such as ideal small-angle X-ray scattering data and experimental solvent accessibility data as the filters to rank the clusters of structurally similar conformations. NAST has been used to predict the 3D structures of the yeast phenylalanine tRNA (76 nt) and the P4-P6 domain of the Tetrahymena thermophile group I intron (158 nt) within 8 Å and 16 Å RMSDs from experimental structures, respectively.[82]

*iFold*. The iFold is a novel web-based algorithm was developed by the Doknolyan's group, which can be used to predict RNA 3D structures from sequences.[83,84] To rapidly explore the possible conformational space of RNA molecules, the model employs a CG representation of three beads per nucleotide and the efficient discrete molecular dynamics simulations with stepwise potentials including base-pairing and base-stacking. The power of iFold has been demonstrated by predicting the 3D structures of 150 RNAs with diverse sequences and the majority of the predicted structures have <4 Å deviations from experimental structures.[83]

*Vfold*. Based on the experimental thermodynamic parameters for helices and the loop entropy from the random walks of virtual bonds in a diamond lattice, the Vfold provides the free energy landscape by enumerating all of the secondary structures including pseudoknots.[85,86] The corresponding 3D structures of the minimum and the local minima in the energy landscape can be built through the fragment assembly based on the secondary structures, and then refined by AMBER[65] energy minimization. The notable advantage of Vfold is the ability to generate RNA structures with cross-linked loops and helices such as pseudoknots and complexes.[86]

*Five-bead Model*. For higher resolution, a five-bead model was developed by Ren and co-workers.[87,88] In the model, each nucleotide is represented by five pseudoatoms, two of which are for the backbone and three are for the base to preferably capture base stacking and pairing. With molecular dynamics simulations and simulated annealing algorithm, the model can be used to predict 3D structures of small RNAs and to capture the 3D structures of large-size RNA with integrating limited experimental data.[88] Since the model is designed to be compatible with existing all-atomistic modeling packages, the predicted CG structures can be directly mapped to all-atomistic structures and refined with all-atomistic force field.[88]

*HiRE-RNA*. The HiRE-RNA is a generic high-resolution CG model with four beads for the RNA backbone, which enables a better description of protein-RNA binding interfaces.[89,90] The base pairing is modeled by hydrogen bonding interactions consisting of 2-, 3- and 4-body terms and the involved interactions include canonical Watson-Crick and wobble base pairs as well as the relatively rare A-C, A-G and U-C pairs. The model has been built in molecular dynamics simulations to study the stability of RNA hairpins[89] and duplexes[90]. However, the parameters of the model may need further adjustment to match the predicted thermodynamic properties of RNAs to the experimental data.

### 3.2.3. A CG model for 3D structure, stability and salt effect

As discussed above, most existing models focus on the 3D structure prediction. Since RNAs are highly charged polyanionic polymers, the RNA structures are sensitive to temperature and ion conditions.[93-96] Is there a model that can predict the RNA structures at a given temperature and ionic condition? To address this issue, a new three-bead CG model (Shi et al. 2014, in submission) has been developed very recently with three beads placed on the existing atoms P, C4' and N9 for the purine (or N1 for the pyrimidine), respectively. With the Monte Carlo simulated annealing algorithm, the model could not only predict 3D structures, but also give reliable predictions on the stability and salt effect for small RNAs. The implicit-solvent/salt force field used in the model contains the local geometric constraints of RNAs as well as the sequence-dependent base pairing and base stacking potentials and electrostatic interactions between phosphate groups.

The model was first validated by predicting the 3D structures of 46 RNAs ($\leq$ 45 nt) from their sequences including pseudoknots and hairpins, with and without bulge/internal loops. For each sequence, from an initial random chain, the RNA folds into a series of native-like structures with the decrease of temperature to a target temperature in the Monte Carlo annealing process. The overall mean RMSD and the overall minimum RMSD between the predicted structures of 46 tested RNAs and the corresponding experimental structures are 3.5 Å and 1.9 Å, respectively. Beyond that, the model can also make reliable predictions on the stability of various RNA hairpins such as melting temperatures with the mean errors 1.0℃ from extensive experimental data at extensive $Na^+$ concentrations[96]. Meanwhile, it can provide the ensemble of probable 3D structures of RNA hairpins at different temperature/salt conditions. Therefore, the model may become a basis for a possible unified model for predicting 3D structures, as well as stability and salt effect, for large RNAs with complex structures.

## 4. Conclusion and perspective

Since understanding RNA structure especially in 3D, is crucial to the understanding of the mysterious RNA world, various computational models have been developed in modeling RNA structures in recent years. RNA secondary structure can be relatively less difficult to predict from sequence alignment, thermodynamics-based dynamic programming algorithms or a combination of approaches. The prediction accuracy on secondary structures would be improved with the increase of the number of the structures deposited in database or with the improvements of the thermodynamic parameters from experiments. However, the art of predicting RNA 3D structure is just beginning to develop and faces challenges.

First of all, predicting 3D structures for large-size RNAs is still a primary challenge, especially the determination of long-range tertiary interactions within RNAs.[97] Some knowledge-based methods can predict the structures with no size limitation, but their requirements for descriptions of known structures and for hands-on interaction with expert user limit their wider application. For physics-based models, one possible way to treat large-size RNAs is to reduce the sampling conformational space. Since RNA folding is generally hierarchical, the secondary structure information (e.g., base-pairing and base-stacking) obtained from the various algorithms and experiments can be used as the constraints to reduce sampling freedom for 3D structure prediction. The architects of some models have taken the lead in assembling the 3D fragments into all-atomistic structures from secondary structures and have been proven to be effective.[98] Moreover, very recent studies show that the use of experimental data such as low-resolution SAXS data and multiplexed hydroxyl radical cleavage analysis can dramatically improve the efficiency and accuracy of structure prediction.[99-103] However, how to effectively and accurately predict 3D structures for large RNAs from sequences is still puzzled. In addition, "coarse-graining", i.e., treating a group of functional atoms as a single bead, is another way to reduce sample space. Generally, to obtain structures at higher resolution, all-atomic structures can be reconstructed from CG structures and optimized with all-atomistic force fields such as AMBER[65] and CHARMM[67].

The second big challenge at the moment is that current studies of RNA functions suggest that the actual functional structures of RNAs might be differ from the theoretical structures predicted on the basis of minimum free energy.[48,104-106] For instance, riboswitches usually perform their functions by changing structure rather than retaining a static native structure.[106] This indicates that the kinetic RNA structures that developed during the folding process can be important for RNA functions.

Although a few methods such as iFold[83] and Vfold[86] can partly predict the 3D structures of the local minima in the energy landscape, it is still difficult for them to provide the folding kinetics of 3D structures because of the incomputable conformations. Nevertheless, it is plausible that one could predict the folding kinetics of a given RNA at secondary structure level using the existing algorithms such as Kinwalker[38] and MPGAfold[39], and then transform the kinetically important structures to 3D representation by the fragment assembly or other physics-based methods to approximately obtain the kinetic RNA folding process at the 3D level, including folding pathways and rates, and kinetic structures.

Thirdly, RNA 3D structures are very sensitive to the solution environment, e.g., temperature, ions, ligands and other macromolecules, due to the structure's high flexibility and the high density of negative charges on RNAs.[8,93,107-116] However, most of the current structure prediction models seldom consider the conditions departing from the room/body temperature and high salt (namely, 1M NaCl). It is a challenge to predict RNA 3D structures at arbitrary temperature/ion conditions. Very recently, a new physics-based CG model have been developed to predict the RNA 3D structures at different salt/temperature conditions from sequences (Shi et al. 2014, in submission), but the model is only applied to small RNAs ($\leq$ 45 nt) and it is not suitable for the solutions with multivalent ions such as $Mg^{2+}$, which have been shown to play very important roles in RNA tertiary structures and functions.[93,116,117] The model can be possibly extended to involve explicit ions while the conformation searching cost would increase correspondingly.[95,118] Furthermore, it is known that cells contain up to 30% by volume of macromolecular species such as proteins, DNA, and RNA.[113] Recent experimental and theoretical studies have suggested that macromolecular crowding can greatly influence the structure stability and kinetic properties of RNAs.[115] Unfortunately, the existing models seldom consider the influence of the macromolecular crowding in RNA 3D structure prediction. Additionally, some RNAs become functional only upon forming complexes with proteins or other molecules. For example, RNA–protein complexing plays an important role in protein synthesis, viral replication, cellular defense and developmental regulation.[119-122] To form complexes, RNAs usually adopt specific structures to provide binding sites for proteins and other molecules.[120-122] However, to model the complex structures of RNAs with other molecules is more challenging, due to the fewer available experimentally discovered structures[121] and the complicated interactions between RNAs and other molecules.

Nevertheless, the recent successes in RNA 3D structure modeling suggest that we can expect many exciting developments in RNA structure prediction in the coming decade.

## Acknowledgments

We are grateful to Shi-Jie Chen, Wenbing Zhang, and Song Cao for valuable discussions.

**Figure and Tables.**

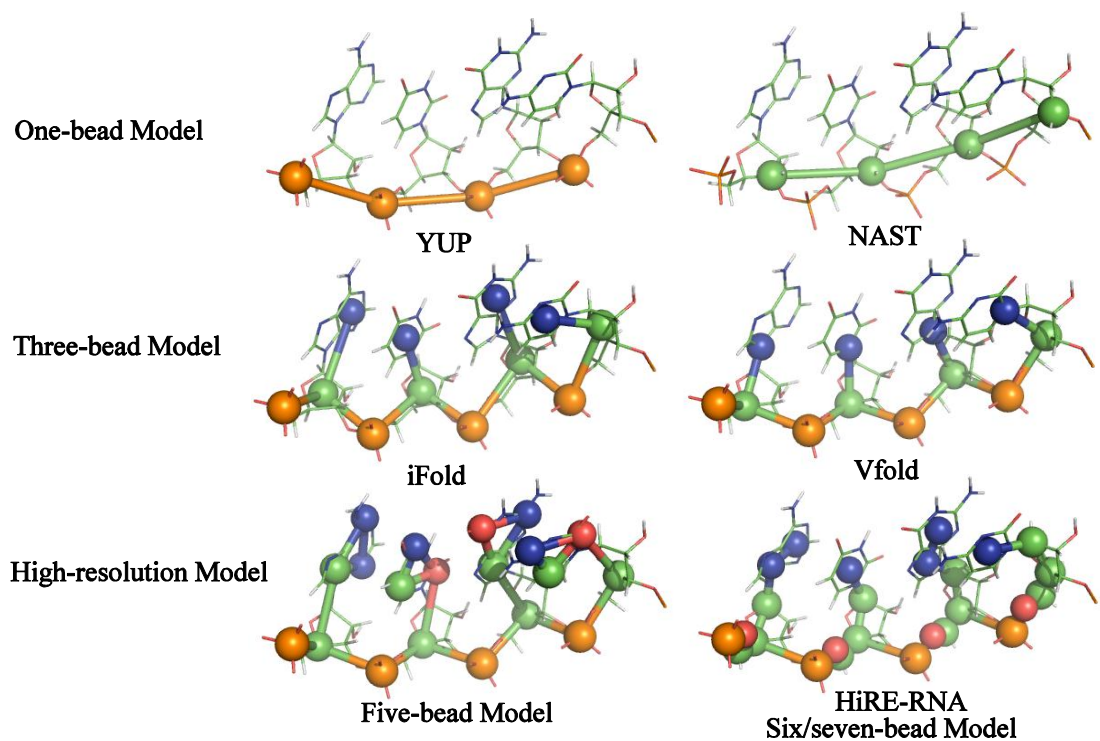

Figure 1. The coarse-grained representations (ball-sticks) of different resolution models for one fragment of sequence: 5'-AUGC-3' superposed on an all-atom representation (thin sticks).

Table 1. The major algorithms for RNA secondary structure prediction[a].

| Algorithms | Refs. | Features | Web page |
|---|---|---|---|
| RNAalifold | [20] | Compute minimum energy structures from aligned sequences using covariation analysis. | http://rna.tbi.univie.ac.at/cgi-bin/RNAalifold.cgi |
| Pfold/PPfold | [21,22] | Predict RNA secondary structures using phylogeny and auxiliary data. | http://daimi.au.dk/~compbio/pfold/ |
| MARNA | [25] | An approach for multiple alignments of RNAs with sequence/structure comparisons. | http://rna.informatik.uni-freiburg.de/MARNA/Input.jsp |
| Mfold | [29] | Search minimization free energy structures by dynamic programming algorithm. | http://mfold.rna.albany.edu/?q=mfold |
| RNAStructure | [30] | Predict RNA secondary structures as well as base pairs probabilities. | http://rna.urmc.rochester.edu/RNAstructureWeb/ |
| RNAfold | [31] | Predict minimum free energy structures and base pair probabilities from single sequences. | http://rna.tbi.univie.ac.at/cgi-bin/RNAfold.cgi |
| Contrafold | [33] | Predict structures using discriminative training and feature-rich scoring. | http://contra.stanford.edu/contrafold/ |
| Sfold | [34] | Predict structures by a statistical sampling based on Boltzmann-weighted ensemble. | http://sfold.wadsworth.org/cgi-bin/index.pl |
| RNAShapes | [35] | Select suboptimal shapes using a reduced representation of RNA structures. | http://bibiserv.techfak.uni-bielefeld.de/rnashapes/ |
| Kinwalker | [38] | A heuristic approach to kinetic RNA folding. | http://www.bioinf.uni-leipzig.de/Software/Kinwalker/ |
| MPGAfold | [39] | Predict folding pathways and functional substructure with genetic algorithm. | http://www-lmmb.ncifcrf.gov/users/bshapiro//mpgaFold/mpgaFold.html |
| Pknots | [41] | Pseudoknot structure prediction using a dynamic programming algorithm. | http://selab.janelia.org/software.html |
| ILM | [42] | An iterated loop matching algorithm for pseudoknots structure prediction. | http://www.cs.wustl.edu/~zhang/projects/rna/ilm/ |
| Hotknot | [43] | A heuristic algorithm for predicting RNA secondary structures including pseudoknots. | http://www.cs.ubc.ca/labs/beta/Software/HotKnots |
| PknotsRG | [46] | Fold medium-sized pseudoknots based on thermodynamics model. | http://bibiserv.techfak.uni-bielefeld.de/pknotsrg/ |

[a] This table summarizes the major algorithms in secondary structure prediction, and provides a URL for downloading or using online for each algorithm.

Table 2. The major algorithms for RNA 3D structure prediction.[a]

| Algorithms | Refs. | Classification | Model | Web page |
|---|---|---|---|---|
| MANIP | [52] | Graphics-based | All-atomic | http://www-ibmc.u-strasbg.fr/upr9002/westhof/index.html |
| ERNA-3D | [53] | Graphics-based | All-atomic | http://owww.molgen.mpg.de/~ag_ribo/ag_brimacombe/ERNA3D/ERNA-3D.html |
| RNA2D3D | [54] | Graphics-based | All-atomic | http://www.ccrnp.ncifcrf.gov/users/bshapiro/software.html |
| S2S/Assemble | [55,56] | Graphics-based | All-atomic | http://bioinformatics.org/assemble/ |
| ModeRNA | [59] | Homology-based | All-atomic | http://iimcb.genesilico.pl/moderna/ |
| RNABuilder | [60] | Homology-based | All-atomic | https://simtk.org/home/rnatoolbox |
| 3dRNA | [61] | Homology-based | All-atomic | http://biophy.hust.edu.cn/3dRNA/3dRNA.html |
| RNAComposer | [63] | Homology-based | All-atomic | http://rnacomposer.ibch.poznan.pl |
| MC-fold/MC-Sym | [68] | Physics-based | All-atomic | http://www.major.iric.ca/MajorLabEn/MC-Tools.html |
| FARNA/FARFAR | [69,70] | Physics-based | All-atomic | http://rosie.rosettacommons.org/ |
| RSIM | [71] | Physics-based | All-atomic | http://www.github.com/jpbida/rsim |
| BARNACLE | [72] | Physics-based | All-atomic | http://sourceforge.net/projects/barnacle-rna/ |
| RNAnbnds | [73] | Physics-based | All-atomic | http://biophy.nju.edu.cn/index-en.htm |
| YUP | [81] | Physics-based | Coarse-grained: One-bead | http://www.harvey.gatech.edu/YammpWeb/default.html |
| NAST | [82] | Physics-based | Coarse-grained: One-bead | https://simtk.org/home/nast |
| iFold | [83,84] | Physics-based | Coarse-grained: Three-bead | http://danger.med.unc.edu/tools.php |
| Vfold | [85,86] | Physics-based | Coarse-grained: Three-bead | http://vfold.missouri.edu/chen-software02.html |
| Five-bead Model | [87,88] | Physics-based | Coarse-grained: Five-bead | http://biomol.bme.utexas.edu/lab/ |
| HiRE-RNA | [89,90] | Physics-based | Coarse-grained: Six/seven-bead | http://www-lbt.ibpc.fr/LBT/index.php?page=lbt&hl=en |

[a] This table summarizes the major methods in RNA 3D structure prediction, and provides a URL for downloading or using online for each method.